\documentclass[12pt,preprint]{aastex}

\newcommand{\kms}          {\mbox{${\rm km~s^{-1}}$}}

\newcommand{\e}            {\mbox{$^{-1}$}}
\newcommand{\ee}           {\mbox{$^{-2}$}}

\def\cm2{\mbox{${\rm cm^{-2}}$}}
\def\h2{\mbox{${\rm H}_2$}}
\def\nh2{\mbox{$n_{\rm H_2}$}}
\def\Nh2{\mbox{$N_{{\rm H}_2}$}}
\def\Mh2{\mbox{$M_{{\rm H}_2}$}}

\def\Mearth{\mbox{$M_\oplus$}}

\def\micron{\hbox{$\mu$m}}
\def\simgt{\lower.5ex\hbox{$\; \buildrel > \over \sim \;$}}
\def\simlt{\lower.5ex\hbox{$\; \buildrel < \over \sim \;$}}

\def\t1c{\mbox{$\theta^1$\,Ori\,C}}
\def\aj{AJ}
\def\apj{ApJ}
\def\apjl{ApJ}
\def\13co{$^{13}$CO}
\def\c18o{C$^{18}$O}
\def\H2{H$_2$}

\def\Mdust{\mbox{$M_{\rm dust}$}}
\def\Tdust{\mbox{$T_{\rm dust}$}}

\def\startfigcap{\vspace*{2.0\baselineskip}\bgroup\leftskip 0.45in\rightskip 0.45in\small}
\def\endfigcap{\par\egroup\vspace*{2.0\baselineskip}}
\def\startfigcapside{\vspace*{2.0\baselineskip}\bgroup\leftskip 4.5in\rightskip 0.4in\small}
\def\endfigcapside{\par\egroup\vspace*{2.0\baselineskip}}
\def\plotfiddle#1#2#3#4#5#6#7{\centering \leavevmode
\vbox to#2{\rule{0pt}{#2}}
\includegraphics{#1}}

\begin{document}

\title{The Dust Properties of Eight Debris Disk Candidates as Determined
by Submillimeter Photometry}
\author{Jonathan P. Williams and Sean M. Andrews}
\affil{Institute for Astronomy, University of Hawaii, 2680 Woodlawn Drive, Honolulu, HI 96822}
\email{jpw@ifa.hawaii.edu,andrews@ifa.hawaii.edu}

\shorttitle{Dust Properties of Debris Disks}
\shortauthors{Williams \& Andrews}

\begin{abstract}
The nature of far-infrared dust emission toward main sequence stars,
whether interstellar or circumstellar, can be deduced from
submillimeter photometry. We present JCMT/SCUBA
flux measurements at 850\,\micron\ toward 8 stars with large photospheric
excesses at 60--100\,\micron. 5 sources were detected at $3\sigma$ or
greater significance and one was marginally detected at $2.5\sigma$.
The inferred dust masses and temperatures range from 0.033 to $0.24\,M_\oplus$
and $43-65$\,K respectively. The frequency behavior of the opacity,
$\tau_\nu\propto\nu^\beta$, is relatively shallow, $\beta < 1$.
These dust properties are characteristic of circumstellar material,
most likely the debris from planetesimal collisions.
The 2 non-detections have lower temperatures, $35-38$\,K,
and steeper opacity indices, $\beta > 1.5$,
that are more typical of interstellar cirrus.
The confirmed disks all have inferred diameters $\simgt 2''$
most lie near the upper envelope of the debris disk mass distribution,
and 4 are bright enough to be feasible for high resolution imaging.
\end{abstract}

\section{Introduction}
Dust around main sequence stars with ages $\simgt 10$\,Myr
must have been formed {\it in situ} due to the short timescales
for collisional destruction and radiative dispersal (Artymowicz 1988).
The micron and sub-micron sized particles
that are detected in infrared images of scattered light
(e.g., Liu 2004) and emission (e.g., Su et al. 2005)
are created from the collisions of larger particles,
possibly from the late stages of planet formation.
The study of such debris disks, therefore, constrain
planet formation models (Kenyon \& Bromley 2004)
and new detections can signpost good targets for imaging of
extrasolar planets (Zuckerman \& Song 2004).

The faint emission from the cold dust is generally only
apparent above the stellar photosphere at wavelengths longer
than about 20\,\micron. The IRAS all sky survey has been the
primary resource for debris disk searches, subsequently augmented
with more targeted surveys by ISO (see de Muizon 2005)
and most recently by Spitzer (Meyer et al. 2004; Rieke et al. 2005).
These instruments are able to define the spectral energy
distribution (SED) of the dust out to near its peak at
about 100\,\micron, but photometry at longer wavelengths
where the emission is optically thin is an essential complement
by providing a measure of the dust mass.
Moreover, by constraining the Rayleigh-Jeans side of the
SED, the disk temperature and luminosity and the
the power law index of the dust opacity can be determined.
These are significantly different in debris disks and the
interstellar medium and thus submillimeter observations
can confirm or reject debris disk candidates that are based
solely on unresolved far-infrared excesses.

Observations of debris disks at long wavelengths are difficult
due to the intrinsic weakness of the emission and the poor
transparency of the atmosphere.
The Submillimeter Common-User Bolometer Array (SCUBA)
on the James Clerk Maxwell Telescope (JCMT)
has been the workhorse instrument to date
(e.g., Holland et al. 1998)
but has still had limited success: less than 20 systems have
been detected at submillimeter wavelengths.
Consequently, the evolution of mass and inner radius,
derived from the dust temperature, is not well understood
(Najita \& Williams 2005).

Far-infrared surveys provide a good starting point
to define a target list for submillimeter observations,
although the coldest disks may be missed
(Wyatt, Dent, \& Greaves 2003; Najita \& Williams 2005).
A recent list of 58 nearby ($< 100$\,pc) debris
disk candidates was published by Zuckerman \& Song (2004)
based on an extensive search of the IRAS Faint Source Catalog
by Silverstone (2000).
Most of their candidates have not been observed beyond 100\,\micron\
but many have large excesses and their SEDs extrapolate to
detectable fluxes in the submillimeter regime.

A program to obtain 850\,\micron\ photometry of stars in the
Zuckerman \& Song catalog began in Spring 2004 with the intent
to survey as many sources with large predicted fluxes as possible.
The unfortunate retirement
of SCUBA in the summer of 2005 ended the project after only 8 sources
had been observed. Nevertheless, the truncated survey yielded
a very high (75\%) detection rate and significant constraints on
the nature of the dust toward the 2 other sources.
This short paper describes the observations in \S2,
the spectral energy distributions of the 8 candidate debris disks
in \S3, discusses the results and implications in \S4,
and concludes in \S5.

\section{Observations}
Photometry observations were conducted using the SCUBA bolometer
at the JCMT on Mauna Kea, Hawaii between March 2004 and July 2005.
For most of the data presented here, the precipitable water vapor
level was between 1 and 2~mm corresponding to zenith optical depths
at 850\,\micron\ between about 0.2 to 0.3.
The optical depth at 450\,\micron\ was generally greater than 1
so high sensitivity measurements were not possible with the
short wavelength bolometers.

Each source, whose position was taken from the 2MASS Point Source Catalog,
was observed for between 50 and 70 minutes with the exception of
HD\,206893 which was observed for 140 minutes.
The pointing was checked via observations of bright quasars near
each source after each major slew and the focus was adjusted
every three hours on average, more frequently at times near
sunrise and sunset. Calibration was performed by observations
of Uranus, Mars and standard sources,
CRL\,618, CRL\,2688, IRC+10216. Based on the agreement of the measured
fluxes with those predicted for these calibrators, we estimate
the photometry is accurate to within 10\% at 850\,\micron.

CO(3--2) emission was detected toward HD\,218396 using the B3 dual
polarization receiver on the JCMT during observations in June 2005.
The emission was relatively strong and a small map was made in the
vicinity of the source to measure its extent.
The map was made by scanning across the field
alternately in right ascension and declination
with a fixed off position $3'$ east of the map center.
The system temperatures decreased from 620\,K to 430\,K as
the source rose from $37^\circ$ to $64^\circ$.
First order baselines were removed and the data gridded and
coadded to produce a fully sampled map with size $70''\times 70''$
and rms 0.2\,K per 0.15\,\kms\ channel. Due to the time taken to
make this map, none of the other sources were observed with the
heterodyne receivers.

\section{Results}
\subsection{Spectral Energy Distributions}
The 8 sources that were observed with SCUBA are listed in Table 1.
The age estimates are from a recent re-evaluation by Mo{\'o}r et al. (2006)
whose target list of 60 debris disk candidates with high
fractional luminosity overlap with 6 of the sources here.
The ages for the two sources not in Mo{\'o}r et al.,
HD\,14055 and HD\,56099, are taken from Zuckerman \& Song (2004).
Distances are derived from Hipparcos parallaxes.
The SCUBA flux measurements and $1\sigma$ statistical
uncertainty are listed in the last column of the table.
A 10\% calibration uncertainty applies to both the
flux and error.

Five of the eight sources, HD\,14055, HD\,15115, HD\,21997, HD\,127821,
and HD\,219396 were detected at 850\,\micron\ at $3\sigma$ significance
or higher. An additional source, HD\,206893, was detected
with a signal-to-noise ratio of 2.5.
It was observed on two consecutive days with consistent
($\sim 2\sigma$) results and is considered a marginal detection.
The SEDs of these six sources are shown in Figure\,1.
The Hipparcos and 2MASS catalogs were used to obtain
the photometry at BV and JHK bands respectively and
Kurucz models appropriate for each stellar type are shown
normalized to the near-infrared data.
Fluxes at mid- and far-infrared wavelengths are from the
IRAS Faint Source Catalog, color corrected for the temperature
of the star at 12 and 25\,\micron\ and for the temperature
of the dust at 60 and 100\,\micron\ where appropriate.
Since the corrections change the SED fit slightly
they are determined iteratively along with the dust temperature.
Additional data points from ISO and Spitzer observations
at $10-90$\,\micron\ from Chen et al. (2006) and
Mo{\'o}r et al. (2006) were added where available.

We model the excess dust emission above the stellar photosphere
with a single temperature greybody,
$F_\nu=B_\nu(\Tdust)(1-e^{-\tau_\nu})\Delta\Omega$.
Here $B_\nu$ is the Planck function,
$\tau_\nu=(\nu/\nu_0)^\beta=(\lambda/\lambda_0)^{-\beta}$
is the optical depth with power law index $\beta$,
and $\Delta\Omega$ is the solid angle.
Because the dust is only detected at $3-6$ wavelengths
in each source, we apply an additional constraint,
$\lambda_0\leq 100$\,\micron. This is based on similar fits to
better sampled SEDs of other debris disks (Dent et al. 2000)
and is also justified a posteriori from the fitted disk parameters below.

The dust excesses were then fit to determine $\beta, \Tdust$, and
$\Delta\Omega$ for a range of values of $\lambda_0$.
The results are shown for a representative source, HD\,15115, in Figure\,2.
Fits to the SEDs of the other sources show a similar behavior:
$\beta$ and \Tdust\ decrease slightly as $\lambda_0$ decreases
but $\Delta\Omega$ increases rapidly.
This is due to the fact that the dust is cold, $\sim 50$\,K,
so the mid- and far-infrared data points all lie on the Wien
side of the greybody where the frequency dependence is dominated by
the exponential term. The optical depth dependence produces only a
slight change to the shape of the fit and, if less than one,
mainly affects the flux scale.

The different fits for the different values of $\lambda_0$
are almost indistinguishable in terms of the least squares
difference and the plotted SED, shown in Figure\,1.
Consequently these data are unable to give an accurate measure
of $\lambda_0$ and therefore also $\Delta\Omega$.
Because $\beta$ and \Tdust\ are only weakly dependent on $\lambda_0$,
however, they are well determined.
The one SCUBA data point is critical in this regard because it lies
on the Rayleigh-Jeans side of the distribution where the emission
is both optically thin (so $\lambda_0$ is only a scaling factor),
and the frequency dependence is a power law and therefore strongly
dependent on $\beta$.

The remaining two sources in the observed sample, HD\,56099 and HD\,78702,
were not detected. Both stars have strong infrared excesses but the
low limit on the 850\,\micron\ flux from the SCUBA observations 
implies a significantly steeper submillimeter spectral index,
or equivalently a higher value for $\beta$,
than determined for the detected sources.
The best three-parameter fit to the observed IRAS and SCUBA
fluxes requires $\beta\geq 2$ for all values of $\lambda_0$.
Mie theory, laboratory measurements, and ISM observations suggest
an upper limit of $\beta=2$ (Pollack et al. 1994; Draine 2006).
The lowest value of $\beta$ consistent with the data is obtained
by fitting the SED with an 850\,\micron\ flux equal to the
$3\sigma$ SCUBA upper limit. The best fit SED is shown for each
source in Figure\,3, and the allowable fits consistent with
the upper limits are overlaid in greyscale.

Table\,2 lists the best fit parameters, \Tdust\ and $\beta$,
and their full range, including errors, for $\lambda_0=3-100$\,\micron.
The derived dust mass and fractional luminosity are listed in
columns 3 and 4 respectively. The dust mass is calculated
from the 850\,\micron\ flux in the standard way and assumes a dust
mass absorption coefficient $\kappa_{850}=1.7$\,cm$^2$\,g\e\
for ease of comparison with other studies.
There is considerable uncertainty over this number
and the masses of the submillimeter emitting particles
may be underestimated by a factor of $3-5$ (see discussion
in Najita \& Williams 2005). This uncertainty dominates that
due to the linear dependence of mass on \Tdust.
The ratio of dust to stellar luminosity
is a measure of the fraction of the starlight absorbed by
the dust and is a good indicator of the likelihood of
detecting scattered light (see \S4). This value is independent
of $\lambda_0$ as the plotted best fit SEDs are almost indistinguishable.
The fact that a single temperature characterizes the emission very
well for each SCUBA detection shows that that the
mass surface density jumps at a specific distance from the
star\footnotemark\footnotetext{
There may still be a significant amount of dust, as measured by area,
in very small dust grains closer to the star. Scattered light images
need not, therefore, show a central hole (Wahhaj et al. 2005).}.
A minimum radius of this inner hole,
$R_{\rm dust}=7.8\times 10^4\,{\rm AU}\,(L_{\rm star}/L_\odot)^{0.5}\,T_{dust}^{-2}$,
is obtained by bounding the dust luminosity by a blackbody
and is listed in column six of Table\,2.

The low masses and large radii that we have determined
allow us to justify, a posteriori, our initial assumption in the SED fits.
Even if the dust were concentrated in a narrow annulus of width
$\Delta R_{\rm dust}=1$\,AU at a radius $R_{\rm dust}$ from the star,
the inferred surface densities for each source are very small,
$\Sigma_{\rm dust}=M_{\rm dust}/2\pi R_{\rm dust}\Delta R_{\rm dust}
\simeq 10^{-3}-10^{-2}$\,g\,cm\ee.
Since $\tau_\lambda\propto\lambda^{-\beta}$,
the wavelength where the disks become optically thin is
$\lambda_0=850\,\micron\,(\kappa_{850}\,\Sigma_{\rm dust})^{1/\beta}
\simeq 1-10$\,\micron.
If the dust were more spread out, or the inner hole larger than
the minimum estimate in Table\,2, the inferred value of $\lambda_0$
would be even smaller.

The five detections and one marginal detection have dust temperatures
ranging from $43-65$\,K and dust opacity index $\beta=0.6-1$.
These are typical values for debris around main sequence stars
(Dent et al. 2000; Najita \& Williams 2005)
and different from the diffuse interstellar medium where
$\Tdust\approx 20$\,K and $\beta\approx 2$ (Boulanger et al. 1996).
Thus these SCUBA observations help to confirm the disk nature
of the far-infrared emission around these stars.

On the other hand, the possible fits to the two non-detections have
lower temperatures, $\Tdust=36-38$\,K, and steeper dust opacity
indices $\beta > 1.5$ than the other sources in the sample.
These values suggest that the IRAS emission is warm interstellar
cirrus (Low et al. 1984) and we conclude that these two sources
are probably not debris disks.

\subsection{Molecular gas along the line of sight to HD\,218396}
A CO 3--2 spectrum was taken toward HD\,218396 and surprisingly strong
emission was detected.
CO disk emission rapidly decreases as the infrared excess decreases
(Dent, Greaves, \& Coulson 2005). No star with such a small
excess has been detected in CO before and the measured integrated
intensity, $I_{\rm CO}=1.6$\,K\,\kms, is more typical of protostellar,
rather than debris, disks.

In fact, it seems unlikely that the CO comes from a disk around the star.
The radial velocity of the gas, $v_{\rm CO}=-5.0$\,\kms\
and star, $v_*=-12.6$\,\kms\ (Mo{\'o}r et al. 2006),
are significantly different and there is no indication that
the star is a spectroscopic binary (Royer et al. 2002).
No significant emission to a $3\sigma$ limit of 0.54\,K per 0.15\,\kms\
channel is detected at the stellar radial velocity.
We also mapped the emission and found it to be extended (Figure\,4).
The CO line probably arises from an unrelated background
high latitude ($b=-35.6^\circ$) cloud along the line of sight,
about $1^\circ$ north and perhaps an extension of cloud 54
in the catalog of Magnani, Blitz, \& Mundy (1985).
The filling factor of high latitude molecular gas is small,
$\approx 3$\%, in the southern hemisphere (Magnani et al. 2000),
but not negligible.

The mid-infrared emission is compact in the Spitzer IRS SL slit,
at $6''$ resolution, but it may possess some extended emission,
containing approximately half the total flux, in the wings of the PSF
in the LL2 slit (Chen, personal communication).
At the much lower IRAS resolution, $\simeq 1'$,
there is no evidence for extended emission.
The SCUBA detection of HD\,218396 has the highest signal-to-noise ratio,
$\simeq 6$, of our sample. There is no emission in the neighboring
bolometers, however, showing that the 850\,\micron\ flux is concentrated
within about $20''$.

Any extended background emission would imply that the measured
fluxes overestimate the disk emission. A uniform scaling
would not change the inferred values of \Tdust\ and $\beta$.
The shallow submillimeter slope, in particular, is strong
evidence that the dust emission is disk-like in nature.
Even if the SCUBA flux were a factor of 2 lower and
the infrared data unchanged, the best fit, $\beta=1$,
is significantly less than ISM dust.

Based on the above, we conclude that HD\,218396 is surrounded by
a debris disk that is detected in the continuum from $30-850$\,\micron,
but not in the CO 3--2 line.

\section{Discussion}
The 8 sources here were selected based on the extrapolation of
far-infrared SEDs to submillimeter wavelengths.
6 of the 8 were detected, although one only at $2.5\sigma$ significance.
This is a far higher success rate than other
submillimeter surveys based solely on young stellar ages,
either via association with massive stars (Wyatt et al. 2003)
or moving group membership (Liu et al. 2004)
and stellar activity (Najita \& Williams 2005).
The stellar age of the 6 detected sources range from $\sim 10$\,Myr
to $\simgt 200$\,Myr and possibly as large as 3\,Gyr.
Rieke et al. (2005) suggest that age is not the only factor in either
disk occurrence or dust mass and that there may be a significant
stochastic influence due to rare collisions of large objects
(see also Song et al. 2005).

Carpenter et al. (2005) showed that the mass of dust particles
traced by millimeter measurements is smaller in debris disks
around main sequence stars than protostellar disks.
Liu et al. (2004) and Najita \& Williams (2005) define an upper envelope
to the debris disk mass distribution that decreases inversely with age.
The new detections here and the revised age estimates in Mo{\'o}r et al. (2006)
do not greatly alter these conclusions (Figure\,5).
4 of the 6 new detections here lie near the upper boundary
of the mass distribution
and two, HD\,127821 and HD\,206893, add to the statistics
in the poorly characterized region where stellar ages
are greater than 300\,Myr.

We have also reexamined the dependence of dust temperature and
minimum inferred radius on the stellar luminosity and age
with the addition of the 6 new sources here.
As in Najita \& Williams (2005), no obvious correlations were found.

The double peak in the SEDs shows that the dust is physically separated
from the stars but the far-infrared and submillimeter observations
discussed here are unable to resolve the disks.
Higher resolution studies are important for measuring disk sizes and
characterizing asymmetries that may signpost planetary companions.
Minimum radii based on the fitted dust temperatures are given in
Table\,2 and range from $1-2''$ in angular size.
These are well within the range of optical and infrared observations
and the issue of detectability is one of contrast.
Two systems, HD\,15115 and HD\,21997, have reasonably high fractional
luminosities, $L_{\rm dust}/L_{\rm star}=6-7\times 10^{-4}$,
that are comparable to AU\,Mic and therefore might be detectable in
scattered light either via ground based coronagraphy
(Kalas, Liu, \& Matthews 2004; Liu 2004) or direct imaging with the
{\it Hubble Space Telescope} (e.g., Krist et al. 2005).

Interferometric imaging at (sub)millimeter wavelengths is
an important alternative means to investigate disk structure
(e.g., Wilner et al. 2002).
The millimeter sized particles that are revealed in such
observations are less affected by radiative forces than
the micron sized particles seen in scattered light.
Consequently, resolved millimeter wavelength images are a more
reliable measure of the gravitational potential of a system
and of any dynamical resonances that might be induced by a planet
(Kuchner \& Holman 2003; Wyatt 2003).
The two systems here, HD\,127821 and HD\,218396, with 850\,\micron\
fluxes greater than 10\,mJy, and diameters $\simgt 3-4''$,
are just accessible to the current
generation of interferometers for high resolution mapping.

\section{Conclusions}
We have carried out 850\,\micron\ photometry of 8 nearby stars
to search for thermal emission from cold circumstellar dust.
With one exception, the rms noise levels of these
SCUBA observations lie between $\sigma=1.5-2.3$\,mJy.
5 sources were detected at $3\sigma$ significance or higher
and an additional one at $2.5\sigma$.
The high detection rate is attributable to their selection
based on high IRAS 60 and 100\,\micron\ excesses.

These 6 sources are a significant addition to the small number
of debris disks detected at submillimeter wavelengths.
The inferred dust masses range from $0.033-0.23\,M_\oplus$.
Several are some of the most massive disks known for their
age and show that even the upper end of the disk mass
distribution may not have been fully explored.
Dust temperatures range from $43-65$\,K, and are similar to other
debris disks. There is no correlation of either temperature of
equivalent black-body radius with stellar luminosity or age.

Far-infrared and submillimeter observations are generally unable
to resolve the emission from the cold dust.
Flux measurements, or even sensitive upper limits,
on the Rayleigh-Jeans side of the SED
regime constrain the dust opacity index and are important for
distinguishing between a disk or interstellar (most likely cirrus) origin.
The former applies for the 6 detections which have $\beta < 1$
and the latter for the 2 non-detections where the combination of
high far-infrared and low submilllimeter fluxes constrain
$\beta > 1.5$.

The photospheric excesses are well fit by single temperature greybodys
implying that the bulk of the dust, as measured by mass, lies far
from the star, ranging from 42 to 84\,AU for each source
or about $1-2''$ in angular size.
Two of the confirmed disks have high fractional luminosities and
may be detectable in scattered light. An additional 2 are strong
enough to image with the current generation of (sub)millimeter
interferometers. Such high resolution observations may show
clumping in the disk due to planetary induced resonances.

There are undoubtedly more sources in the Zuckerman \& Song catalog,
and perhaps others in the plethora of new Spitzer results
(e.g., Bryden et al. 2006),
that would be detectable if SCUBA were available.
Its successor, SCUBA-2, will be about a factor of 3 times more
sensitive for photometry and far quicker for mapping (Holland et al. 2003).
It will distinguish between cirumstellar and interstellar emission
and provide mass measurements and dust opacities for
perhaps as many as an order of magnitude more debris disks.

\acknowledgments
This work is supported by the NSF through grant AST-0324328.
We thank Christine Chen and Mike Jura for helpful discussions
concerning Spitzer IRS photometry, and Michael Liu and Joan Najita
for comments on the manuscript.
This research has made use of the SIMBAD database and
the Two Micron All Sky Survey, which is a joint project of the
University of Massachusetts and IPAC/Caltech, funded by NASA and NSF.

%\clearpage

\clearpage
\begin{table}
\begin{center}
TABLE 1\\
Sourcelist and submillimeter photometry\\
\vskip 2mm
\begin{tabular}{lrcccc}
\hline\\[-2mm]
Source & HIP &SpT &  Age &  $d$ & $F_{850}$ \\
       &     &     & (Myr)& (pc) &  (mJy)    \\[2mm]
\hline\hline\\[-3mm]
HD\,14055  &  10670 &   A1V  &  100?               & 36.1 &  $5.5\pm 1.8$ \\
HD\,15115  &  11360 &   F2   &  $12^{+8}_{-4}$     & 44.8 &  $4.9\pm 1.6$ \\
HD\,21997  &  16449 & A3IV/V &  $20^{+10}_{-10}$   & 73.8 &  $8.3\pm 2.3$ \\
HD\,56099  &  35457 &   F8   & $>500?$             & 86.8 &  $1.4\pm 1.9$ \\
HD\,78702  &  44923 & A0/A1V & $220^{+100}_{-140}$ & 79.9 &  $0.3\pm 2.3$ \\
HD\,127821 &  70952 &  F4IV  & [200,3400]          & 31.7 & $13.2\pm 3.7$ \\
HD\,206893 & 107412 &  F5V   & $<2800$             & 38.9 &  $3.8\pm 1.5$ \\
HD\,218396 & 114189 &  A5V   & [20-150]            & 39.9 & $10.3\pm 1.8$ \\[2mm]
\hline
\end{tabular}
\end{center}
\end{table}

%\clearpage
\begin{table}
\begin{center}
TABLE 2\\
Dust properties\\
\vskip 2mm
\begin{tabular}{lllrrrl}
\hline\\[-2mm]
Source & \Tdust & $\beta$ & \Mdust & $L_{\rm dust}/L_{\rm star}$ & $R_{\rm dust}$ & Notes \\
 & (K) & & (\Mearth) & $(10^{-4})$ & (AU) & \\[2mm]
\hline\hline\\[-3mm]
HD\,14055  & $65_{-10}^{+4}$ & $0.98^{+0.17}_{-0.30}$ & 0.033  & 1.0  & 84 & \\
HD\,15115  & $62_{-5}^{+3}$  & $0.73^{+0.16}_{-0.27}$ & 0.047  & 5.8  & 34 & \\
HD\,21997  & $56_{-6}^{+1}$  & $0.86^{+0.14}_{-0.27}$ & 0.24   & 7.1  & 75 & \\
HD\,56099  & $38_{-5}$ & $2.00_{-0.51}$ & $<0.091$ & $\sim 11$  & &
   interstellar cirrus? \\
HD\,78702  & $36_{-7}$ & $2.00_{-0.19}$ & $<0.018$ & $\sim 2.6$ & &
   interstellar cirrus? \\
HD\,127821 & $43_{-9}^{+10}$ & $0.64^{+0.44}_{-0.34}$ & 0.096  & 3.0  & 67 & \\
HD\,206893 & $53_{-10}^{+5}$ & $0.82^{+0.25}_{-0.36}$ & 0.033  & 2.8  & 42 &
   marginal detection \\
HD\,218396 & $50_{-3}^{+1}$  & $0.71^{+0.10}_{-0.21}$ & 0.10   & 2.8  & 66 & 
   CO detection \\[2mm]
\hline
\end{tabular}
\end{center}
\end{table}

\clearpage
\begin{figure}[ht]
\plotfiddle{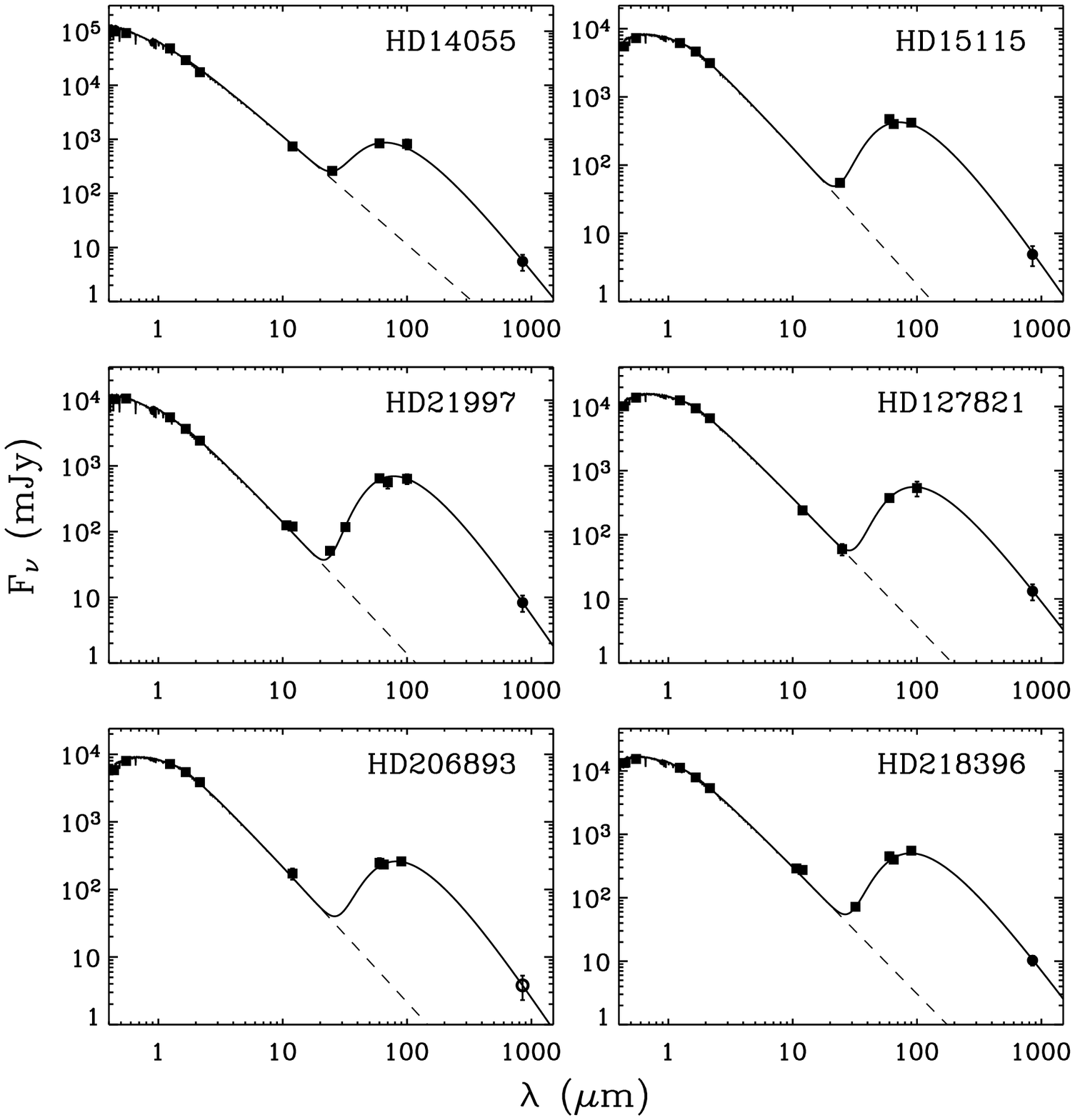}{0pt}{0}{75}{75}{-225}{-540}
\end{figure}
\vskip 5.9in
\noindent{\bf Figure 1:}
%\plotone{f1.eps}
%\vspace*{-35mm}
%\caption{
Spectral energy distributions for the 6 sources with SCUBA detections.
The square symbols show fluxes at BVJHK bands and in the mid- and far-infrared
from $12-100\,\mu$m. The circular symbols show the SCUBA measurements
at $850\,\mu$m, filled where the signal-to-noise ratio is greater than or
equal to 3, and open for the the marginal ($2.5\sigma$)
detection of HD\,206893.
The solid line shows the sum of a Kurucz model of
the stellar photosphere and greybody dust emission.
The latter dominates at $\lambda\simgt 30\,\mu$m and
the extension of the Kurucz model is shown as a dashed line.
Parameters of the greybody fits are shown in Table\,2.
%}

\clearpage
\begin{figure}[ht]
\plotfiddle{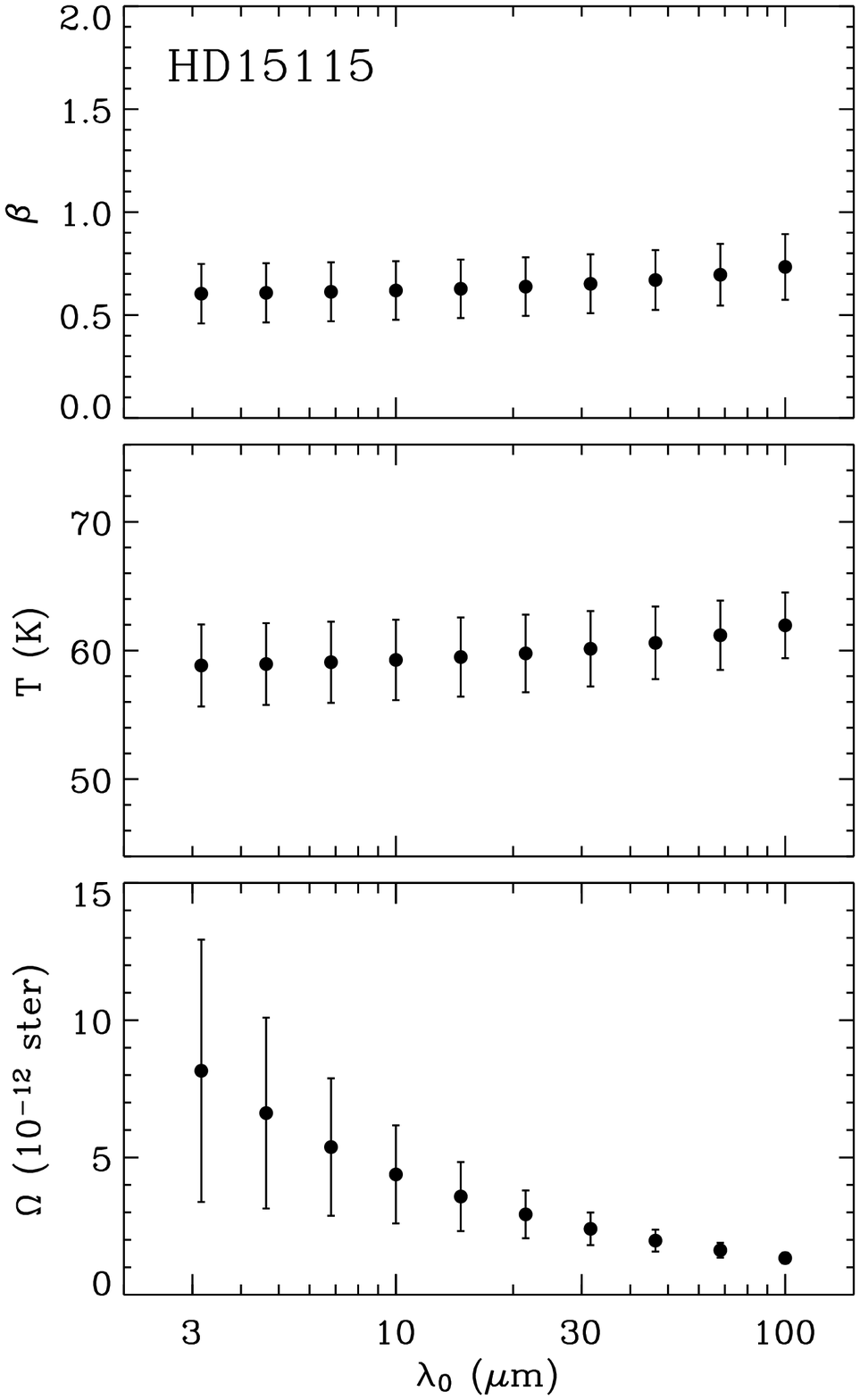}{0pt}{0}{50}{50}{-155}{-360}
\end{figure}
\vskip 4.8in
\noindent{\bf Figure 2:}
%\plotone{f2.eps}
%\vspace*{-20mm}
%\caption{
Variation of the best fit SED parameters with the wavelength,
$\lambda_0$, at which the disk becomes optically thin for HD15115.
The top two panels show that the dust opacity spectral index, $\beta$,
and temperature, $T$, are almost independent of $\lambda_0$
and therefore well constrained.
The lower panel shows that the dust emitting area, $\Omega$, rapidly
increases as $\lambda_0$ decreases and is therefore poorly determined.
Simliar results are found for the SED fits of the other sources.
%}

\clearpage
\begin{figure}[ht]
\plotfiddle{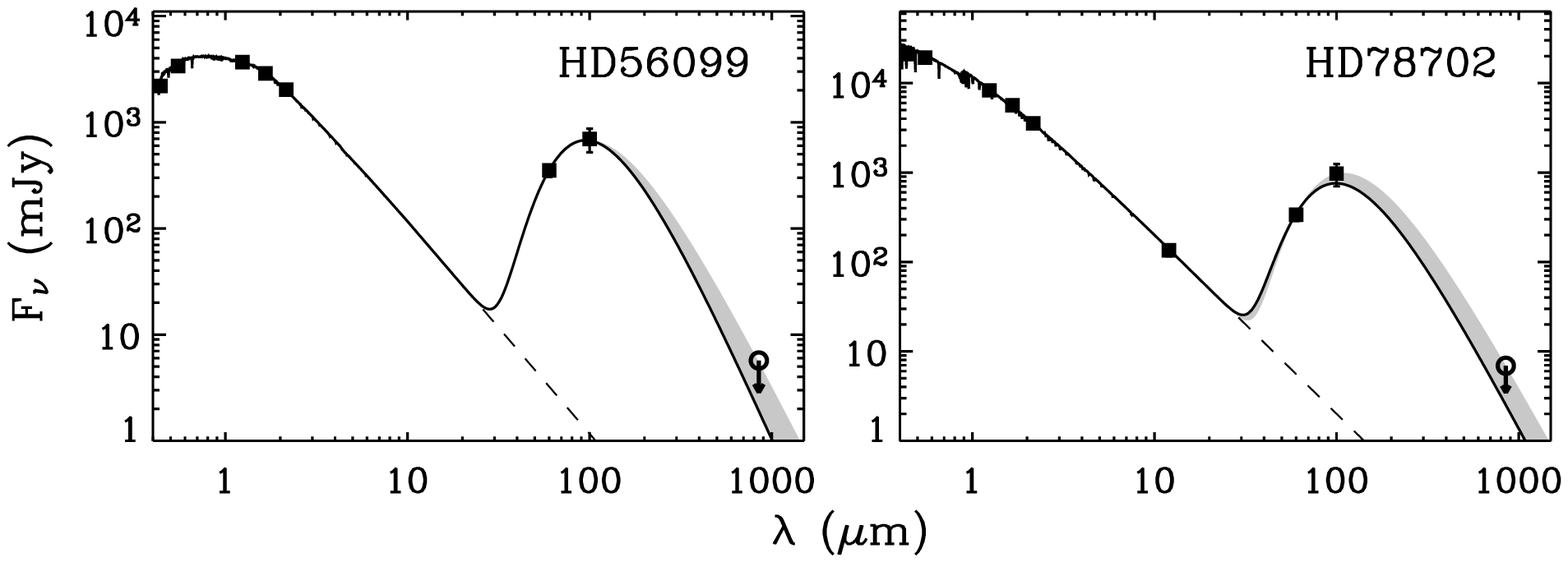}{0pt}{0}{80}{80}{-240}{-580}
\end{figure}
\vskip 2.2in
\noindent{\bf Figure 3:}
%\plotone{f3.eps}
%\vspace*{-85mm}
%\caption{
Spectral energy distributions for the 2 sources that were not
detected by SCUBA. The data and greybody fit to the dust excess
are defined as in Figure\,2.
The grey shaded regions show the allowable fits that are
consistent with the $3\sigma$ upper limit to the $850\,\mu$m flux.
Parameters of the greybody fits are shown in Table\,2.
%}

\clearpage
\begin{figure}[ht]
\plotfiddle{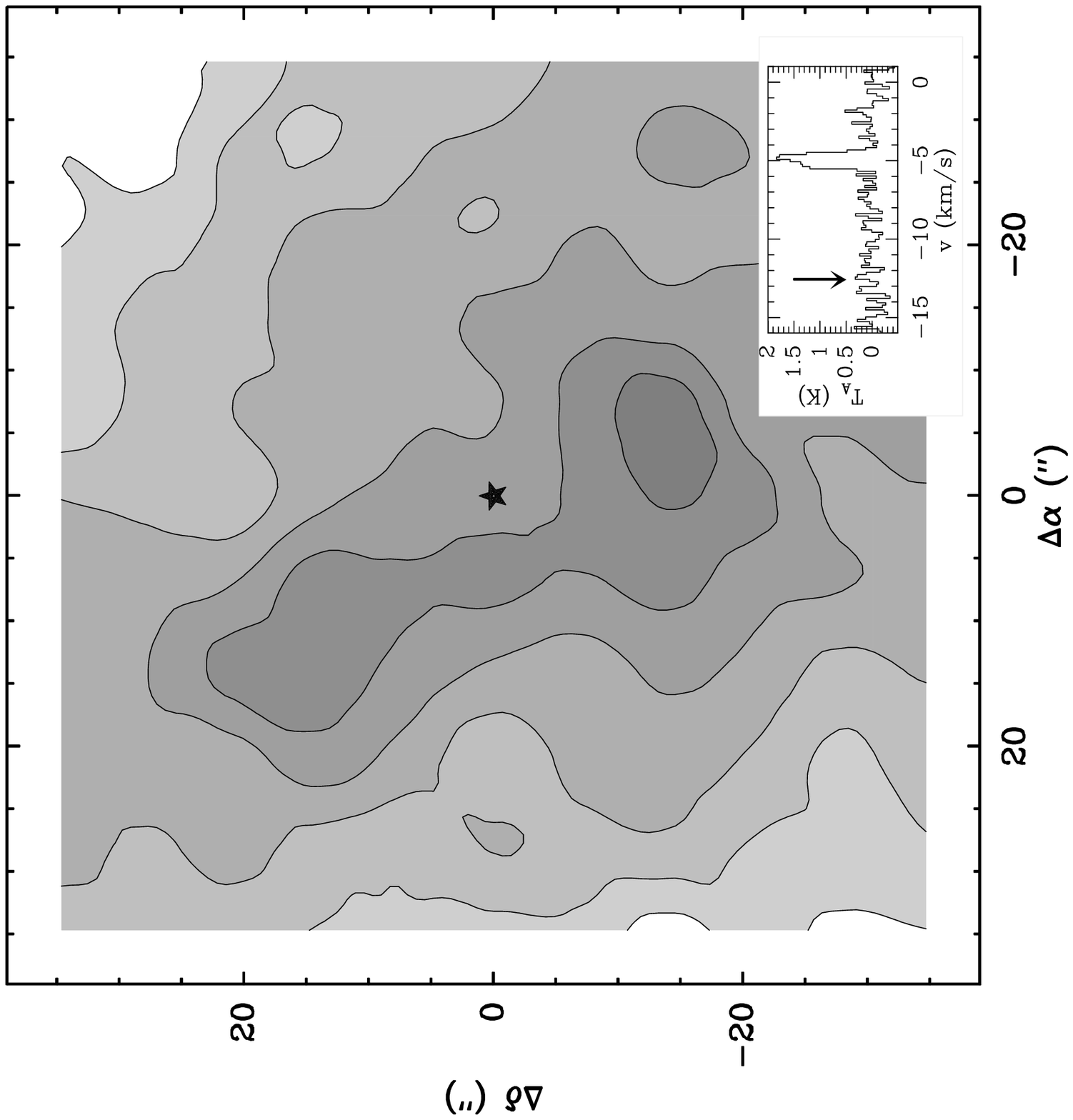}{0pt}{-90}{60}{60}{-250}{35}
\end{figure}
\vskip 4.0in
\noindent{\bf Figure 4:}
%\plotone{f4.eps}
%\caption{
Contours of CO 3--2 emission toward HD\,218396.
The integration range is -6 to -4\,\kms\ and the
contour levels are 0.25\,K\,\kms. The inset shows the spectrum
toward the star and the arrow marks its radial velocity
where any disk emission should be seen.
%}

\clearpage
\begin{figure}[ht]
\plotfiddle{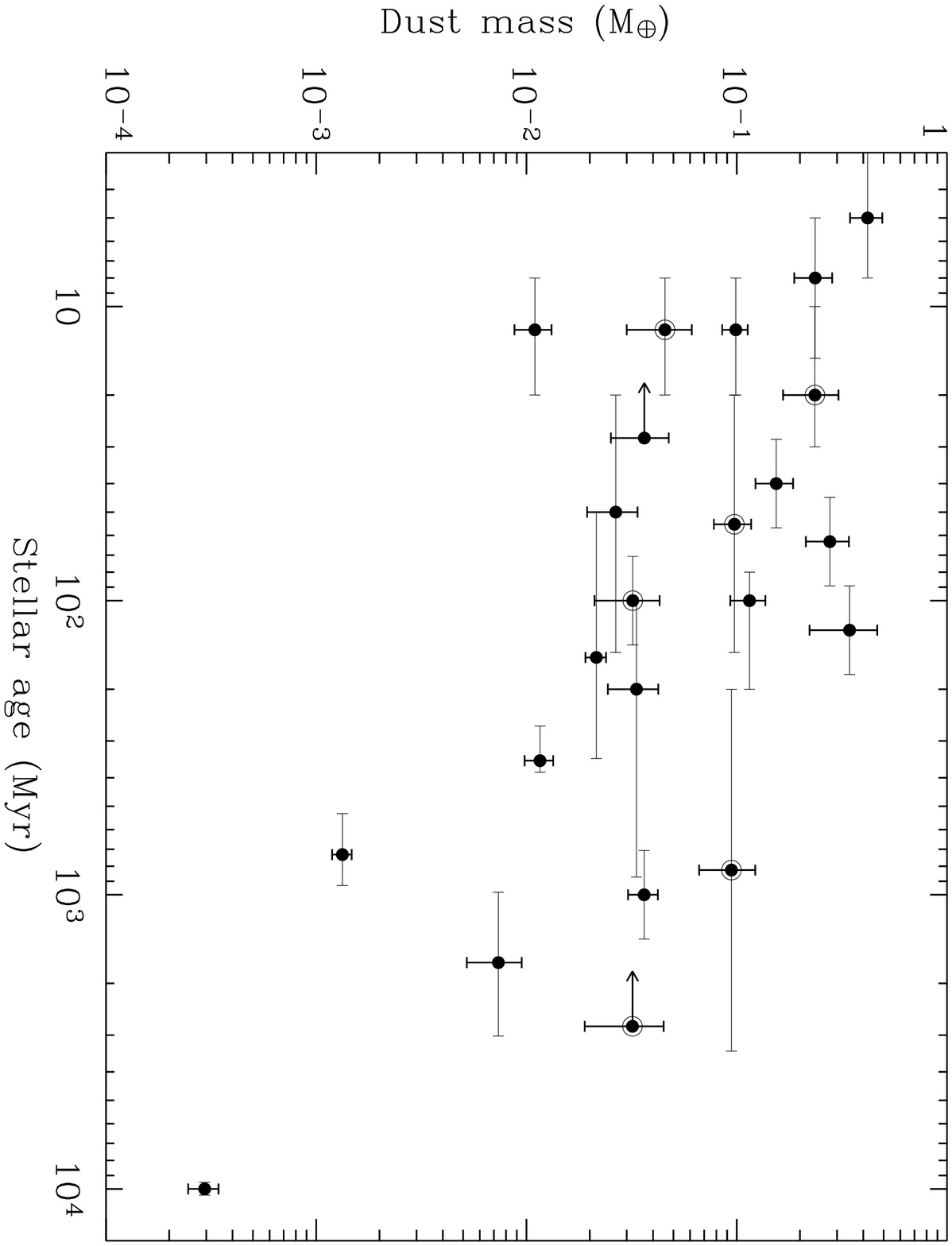}{0pt}{90}{60}{60}{250}{-330}
\end{figure}
\vskip 4.4in
\noindent{\bf Figure 5:}
%\vspace*{-30mm}
%\plotone{f5.eps}
%\vspace*{-20mm}
%\caption{
The dust mass in debris disks versus stellar age.
These points represent the known SCUBA 850\,\micron\ debris disk
detections in the literature
(Wyatt et al. 2003; Sheret et al. 2004; Liu et al. 2004;
Greaves et al. 2004; Wyatt et al. 2005; Najita \& Williams 2005).
They have been converted to a disk mass for the same mass absorption
coefficient, $\kappa_{850}=1.7$\,cm$^2$\,g\e,
and dust temperature given in the SCUBA detection paper.
The error bars in the masses include the stated statistical uncertainty
plus an assumed 10\% calibration uncertainty.
The error bars in the stellar ages are from Mo{\'o}r et al. (2006),
where available, else from the discovery paper.
If no error was given, a minimum and maximum age spanning a factor
of two and centered on the stated age is plotted.
The 6 new measurements presented in this
paper are shown with an outer ring around the central dot.
%}


\begin{references}
%\parskip=0pt
%\bigskip

%\pp Ardila, D. R., Golimowski, D. A., Krist J. E., Clampin, M.,
%        Williams, J. P., Blakeslee, J. P., Ford, H. C., Hartig G. F.,
%        \& G. D. Illingworth, G. D. 2004, ApJ, 617, L147

\reference{}  Artymowicz, P.\ 1988, \apjl, 335, L79

\reference{}  Bryden, G., et al.\ 2006, \apj, 636, 1098

\reference{}  Boulanger, F., Abergel, A., Bernard, J.-P., Burton, W.~B., Desert, F.-X.,
    Hartmann, D., Lagache, G., \& Puget, J.-L.\ 1996, \aap, 312, 256 

\reference{}  Carpenter, J.~M., Wolf, S., Schreyer, K., Launhardt, R., \& Henning, T.\
    2005, \aj, 129, 1049

%\reference{}  Chen, C.~H., et al.\ 2005, \apj, 634, 1372 

\reference{}  Chen, C.~H., et al.\ 2006, ApJS, in press (astro-ph/0605277)

\reference{}  de Muizon, M.~J.\ 2005, Space Science Reviews, 119, 201

%\reference{}  Decin, G., Dominik, C., Waters, L.~B.~F.~M., \& Waelkens, C.\
%    2003, \apj, 598, 636

\reference{}  Dent, W.~R.~F., Walker, H.~J., Holland, W.~S., \& Greaves, J.~S.\
    2000, \mnras, 314, 702

\reference{}  Dent, W.~R.~F., Greaves, J.~S., \& Coulson, I.~M.\ 2005, \mnras, 359, 663

\reference{}  Draine, B.~T.\ 2006, \apj, 636, 1114 

\reference{}  Greaves, J. S., Wyatt, M. C., Holland, W. S., \& Dent, W. R. F. 
        2004, MNRAS, 351, L54

\reference{}  Holland, W.~S., et al.\ 1998, \nat, 392, 788

\reference{}  Holland, W.~S., Duncan, W., Kelly, B.~D., Irwin, K.~D., Walton, A.~J.,
    Ade, P.~A.~R., \& Robson, E.~I.\ 2003, \procspie, 4855, 1

\reference{}  Kalas, P., Liu, M. C., \& Matthews, B. C. 2004, Science, 303, 1990

\reference{}  Krist, J.~E., et al.\ 2005, \aj, 129, 1008

\reference{}  Kenyon, S. J. \& Bromley, B. C. 2004, AJ, 127, 513

\reference{}  Kuchner, M.~J., \& Holman, M.~J.\ 2003, \apj, 588, 1110

\reference{}  Liu, M.~C.\ 2004, Science, 305, 1442

\reference{}  Liu, M.~C., Matthews, B.~C., Williams, J.~P., \& Kalas, P.~G.\
    2004, \apj, 608, 526 

\reference{}  Low, F.~J., et al.\ 1984, \apjl, 278, L19 

\reference{}  Magnani, L., Blitz, L., \& Mundy, L.\ 1985, \apj, 295, 402 

\reference{}  Magnani, L., Hartmann, D., Holcomb, S.~L., Smith, L.~E., \& Thaddeus, P.\
    2000, \apj, 535, 167

\reference{}  Meyer, M.~R., et al.\ 2004, \apjs, 154, 422 

\reference{}  Mo{\'o}r, A., {\'A}brah{\'a}m, P., Derekas, A., Kiss, C., Kiss, L.~L.,
    Apai, D., Grady, C., \& Henning, T.\ 2006, \apj, 644, 525 

\reference{}  Najita, J., \& Williams, J.~P.\ 2005, \apj, 635, 625

\reference{}  Pollack, J.~B., Hollenbach, D., Beckwith, S., Simonelli, D.~P., Roush, T.,
    \& Fong, W.\ 1994, \apj, 421, 615 

\reference{}  Rieke, G.~H., et al.\ 2005, \apj, 620, 1010 

\reference{}  Royer, F., Grenier, S., Baylac, M.-O., G{\'o}mez, A.~E., \& Zorec, J.\
    2002, \aap, 393, 897 

\reference{}  Sheret, I., Dent, W. R. F., Wyatt, M. C. 2004, MNRAS, 348, 1282

\reference{}  Silverstone, M.~D.\ 2000, Ph.D.~Thesis, UCLA

\reference{}  Song, I., Zuckerman, B., Weinberger, A.~J., \& Becklin, E.~E.\ 2005,
    \nat, 436, 363

\reference{}  Su, K.~Y.~L., et al.\ 2005, \apj, 628, 487

\reference{}  Wahhaj, Z., Koerner, D.~W., Backman, D.~E., Werner, M.~W., Serabyn, E.,
    Ressler, M.~E., \& Lis, D.~C.\ 2005, \apj, 618, 385 

\reference{}  Williams, J. P., Najita, J., Liu, M. C., Bottinelli, S.,
    Carpenter, J. M., Hillenbrand, L. A., Meyer, M. R., \& Soderblom, D. R.
    2004, ApJ, 604, 414

\reference{}  Wilner, D. J., Holman, M. J., Kuchner, M. J., \& Ho, P. T. P.
    2002, ApJ, 569, L115

\reference{}  Wyatt, M.~C., Dent, W.~R.~F., \& Greaves, J.~S.\ 2003, \mnras, 342, 876 

\reference{}  Wyatt, M. C., Greaves, J. S., Dent, W. R. F., \& Coulson I. M. 2005, 
        ApJ, 620, 492

\reference{}  Wyatt, M.~C.\ 2003, \apj, 598, 1321 

\reference{}  Zuckerman, B., \& Song, I.\ 2004, \apj, 603, 738 

%\pp Zuckerman, B., Forveille, T., \& Kastner, J. H., 1995, 
%     Nature, 373, 494
\end{references}
\end{document}